\documentclass[12pt]{article}

\usepackage{graphicx}

\begin{document}

\title{Evolving network - simulation study. \\
From regular lattice to scale free network}

\author{Danuta Makowiec\\
Institute of Theoretical Physics and Astrophysics\\
Gda\'nsk University, ul. Wita Stwosza 57, 80-952 Gdansk, Poland \\
 e-mail{\it fizdm@univ.gda.pl}}

\maketitle

\centerline{Abstract}

{\noindent\small
The Watts-Strogatz algorithm of transferring the square lattice to a small world network is modified by introducing preferential rewiring constrained by connectivity demand. The evolution of the network is two-step: sequential preferential rewiring of edges controlled by $p$   and updating the information about changes done. The evolving system self-organizes into stationary states. The topological transition in the graph structure is noticed with respect to $p$. Leafy phase - a graph formed by multiple connected vertices  (graph skeleton) with plenty of leaves attached to each skeleton vertex emerges when $p$ is small enough to pretend asynchronous evolution.  Tangling phase where edges of a graph circulate frequently among low degree vertices occurs when $p$ is large. There exist conditions at which the resulting stationary network ensemble provides networks which degree distribution exhibit  power-law decay in large interval of degrees.
}


\section{\label{sec:Intro} Introduction}
From communication networks like World Wide Web or phone networks, through nets of social relations, namely networks of acquaintances or collaborating scientists, to biological systems where protein networks, neural networks or cell metabolisms are considered --- all of them manifests a similar structural organization: small-world properties (short distances , strong clustering) and the power-law vertex degree distribution. 
See, e.g., \cite{AlbertBarabasi02,DorogovtsevMendes02,Newman03,LNinP} for review of data analysis and bibliography. Most of these natural and man-made networks change in time. Although these networks undergo  different restructuring processes, their crucial statistical properties are time independent. The networks are self-organized to stationary states. Any stationary state can be realized by many (usually huge)  number of microscopically different configurations. Evolution of a stationary state denotes that  microscopic changes performed do not influence the global characteristics.

In the present paper we propose a stochastic microscopic dynamic rule which leads a regular lattice into the stationary network state. By simulations we show the relation between the details of microscopic rule and the distribution of the vertex degree in a stationary state. Our proposition is based the Watts-Strogatz construction of the small-world network by rewiring edges \cite{WattsStrogatz98}.

Here, we do not consider any dynamics on a network. However, our studies are motivated by models of large spatially extended systems with short-range interactions, like Ising models, in which a spin is attached to each  vertex \cite{Ising,Herrero,China,Stauffer,DorogovtsevMendesGoltsev,Dorogovtsev}. Such networks can model the magnetic nanomaterials. Electronic components are represented as vertices  and the wires are edges  \cite{Cancho}. The neural networks may also be considered as kind of electronic circuits \cite{Fernandez,Mathias} following the idea of `save wiring' as an organizing principle of the brain \cite{Cherniak}. 

Our research is aimed on giving hints about the matter in which changes are performed on two levels. The first level means that the network topology evolves. The second level denotes that spins located in nodes are adjusted according to the new network structure.  Preliminary study of such materials  can be found in  \cite{Makowiec04,MakowiecFerro,MakowiecNeuro}. Stochastically evolving connections between at average constant number of vertices can also be viewed as the playground for modeling community structures in networks \cite{NewmanGirvan} or coauthorship networks \cite{Newman}.

The studies of the network topology began with the random graph theory of Erd\"os and R\'enyi \cite{ErdosRenyi}.  The proposition of Watts and Strogatz \cite{WattsStrogatz98} that followed, called small world network, captures  the features of regular lattice and random graph. The algorithm may be summarized as follows: begin with a regular lattice, then rewire an edge with some probability $P$. Traditionally, the evolution in the Watts-Strogatz network is measured by $P$. Let us divide the process of rewiring of $P$ -part of edges into $t$ substeps such that $P= t*p$. Hence, we can say that the $p$- part of edges is rewired synchronously at each time step. If rewiring is  stochastic, like in the Watts-Strogatz algorithm, then the resulting network after $t$ steps with $p$-part edges rewired each time step is equivalent to a network obtained after one step with $P=t * p$ -part edges rewired. However, if  rewiring goes preferentially and information about  modifications introduced is updated once  in a time step then different networks can be observed. 

We show, by simulation, that under some conditions the resulting networks are different from  networks obtained in the 1-step evolution with the corresponding $P$. 
In particular, it appears that if rewiring is accompanied  with  `synchronized' preference then the self-organization of the network state occurs. Namely,  the stable vertex degree distribution is reached though the evolution of edges continuous.
It is said that the ensemble of graphs emerges and the evolution walks on graphs belonging to  this ensemble \cite{Farkas}.

Moreover, we present arguments that the topological transition between the graph ensembles takes place. If $p$ values are small then, so-called, {\it leafy phase} emerges. The phase ensemble consists of  graphs formed by multiple inter connected vertices, called  graph skeleton, with plenty of leaves attached to each skeleton vertex. Occurrence of connections between vertices of similar properties such as, e.g., similar degrees, is termed assortativity and the high probability of connections between vertices with different degrees is termed disassortativity \cite{Farkas}. The strong assortativity between hubs  present in the stationary state results from  preferences in dynamics considered. Moreover, the vertices of the graph skeleton are surrounded by leaves, namely each hub belongs to a star-like subgraph. Hence together with strong assortativity of hubs the strong disassortativity is present also.
On the other hand, if $p$ is large enough then the network stabilizes in, so-called, {\it  tangling phase} where edges of a graph circulate frequently among low degree vertices. 

All network ensembles are characterized by  the distinct degree distributions with exponentially vanishing tails. However, when the parameters driving the network  evolution  are specially  adjusted, then the power-law decay appears in the rather wide interval of vertex degrees. The exponent $\gamma$  for this decay is  close to $2$ what suggests strong inhomogeneity in the network.

The algorithm is presented in Section 2. Section 3 contains results obtained by  simulations. The discussion and  development of the model is proposed in  Section 4.

\section{\label{sec:Alg}Algorithm}

Let us number  vertices of the lattice as  $1,2,\dots, N$. Then each edge is characterized by the two numbers ($from$, $to$) representing two vertices linked by this edge. The graph is represented as the vector of size $N$ of lists of vertices  --- neighbors of subsequent vertices. The graph is not considered as directed, though the algorithm can easily be adapted to a directed one.

There are two basic  parameters in the model:  $p$--- probability to rewire an edge each time step, and   $T$--- threshold in the preference function.

\begin{figure}
\begin{center}
\includegraphics[width=0.6\textwidth]{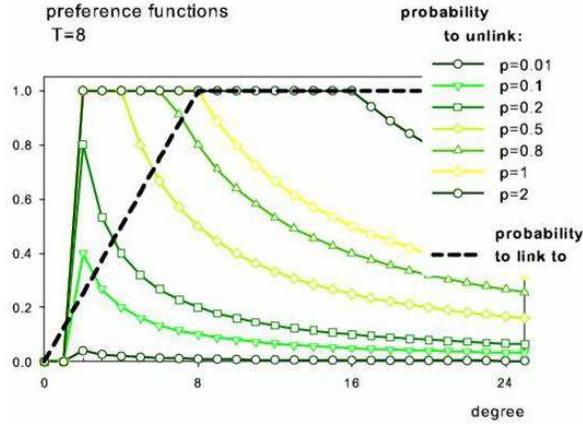} 
\end{center}
\caption{The preferences in  case $T=8$. Notice, that  $p_{ul}(1)=0$ and  $p_{ul}(T)=p$ always, (color on line). }
\label{prefy}
\end{figure}

The synchronized preferential rewiring step means:
\begin{enumerate}
\item Choose at random any of $N$ nodes. For the node $i$ chosen the set of its edges $\{(i,j_i)\}$ is reviewed. The following decisions are made:\\
 (i) the subset $ \{ j_{i^*} \} \subset \{ j_i \} $ of edges to rewire is selected with probability $p_{ul}$ calculated as follows, see Fig.~\ref{prefy}:
$$ p_{ul}(j_{i^*})= 
        {0  \qquad \qquad \qquad \quad{\rm if}\quad {\rm deg}(j_{i^*})=1 \atop
    \min(1,\frac{pT}{{\rm deg}(j_{i^*})})\quad{\rm if}\quad {\rm deg}(j_{i^*})>1}  $$
  
      (ii) for each $j_{i^*}$ selected to rewire the new attachment is assigned $l_{i*}$with probability to link to $l_{i^*}$-node  $p_l$  given by 

 $$ p_{l} (l_{i^*}) = \min(1,\frac{{\rm deg}(l_{i^*})}{T})  $$
  
\item The global information on the vertex degrees is updated.
\end{enumerate}

\begin{figure}
\begin{center}
\includegraphics[width=0.6\textwidth]{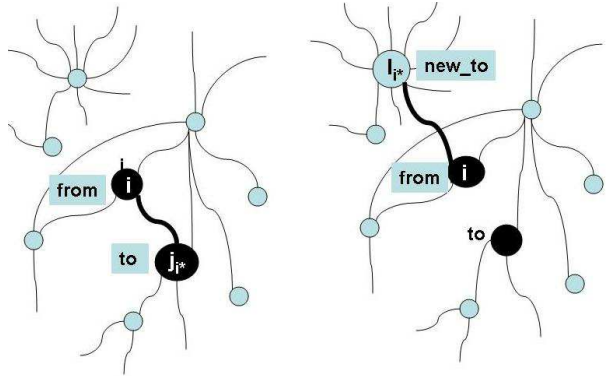}
\end{center}
\caption{\label{pice} An edge linking vertices $from$ and  $to$ (left graph) is rewired to the edge between vertices $from$ and $new\_to$ (right graph)(color on line)}. 
\end{figure}

Since there are possible differences in applying Watts-Strogatz idea of rewiring, below we give the evolution procedure. The basic procedure, called, EdgeEvolution, requires three parameters: $from$, $p$ and $T$. The illustration of the networks {\it before} (left graph) and {\it after}(right graph) rewiring is shown in Fig.~\ref{pice}.

\noindent{\it EdgeEvolution ( $from$, $p$, $T$) :}\\
\leftline{(a)for each vertex $to$ from the list of edges of the vertex}
\leftline{$from$ do}
\leftline{(b)\hspace{0.5cm} if deg$(to)>1 $ then}
\leftline{(c)\hspace{1.0cm}choose $\xi_1\in [0,1]$ at random and }
\leftline{(d)\hspace{1.0cm}if $\xi_1 <\frac{pT}{{\rm deg} (to)}$ then accept the vertex $to$}  \leftline{\hspace{1.3cm} for $unlinking$}
\leftline{(e)for each vertex $to$ accepted for $unlinking$}
\leftline{(f)\hspace{0.5cm}choose $new\_to\in \{ 1,2,\dots, N\}$ at random  but }
\leftline{\hspace{1.0cm}$new\_to \neq from$}
\leftline{(g)\hspace{0.5cm}choose $\xi_2\in [0,1]$ at random  and }
\leftline{(h)\hspace{0.5cm}if $\xi_2 <\frac{{\rm deg} (new\_to) }{T}$
then accept $new\_to$, i.e.,}
$$edge(from,to):= edge(from, new\_to)$$
\leftline{(i)\hspace{0.5cm}otherwise go to (f)}

Remarks:

--- The algorithm conserves both the number of vertices and the number of edges. Steps (a)--(d) prepare the list of edges of the vertex $from$ to be rewired, while steps (f)--(i) fix new connections. Each edge accepted for rewiring must be rewired (see the loop(i)).  

--- The condition (b) is introduced to avoid presence of zero-degree vertices. Any vertex with a degree equal to 1 cannot be unlinked. Otherwise we face the problem of dramatically increasing number of isolated nodes. However, this condition does not protect a graph from being  disconnected. 

--- A randomly selected  new vertex for linking to must be different from the $from$  vertex to avoid loops in a graph, line (f), however, multiple edge links are not forbidden.

--- Since  information regarding the vertex degree is updated  after each time step, hence  if the network is evolving with small $p$, then the evolution may be seen as asynchronous.

\begin{figure}
\begin{center}
\includegraphics[width=0.6\textwidth]{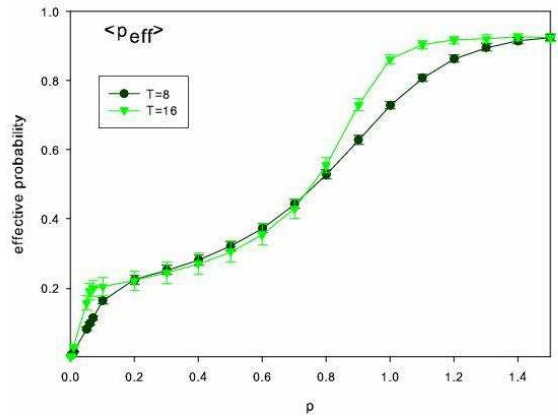}
\end{center}
\caption{Normalized averages of numbers of rewired edges (with standard deviation errors) each time step for different $p$ and $T=8, 16$, (color on line). }
\label{p_eff}
\end{figure}

--- If deg$(to)=T $ then the probability to unlink from the $to$ vertex is equal to $p$ while if deg$(to)<T $ (deg$(to)>T $) then probability to unlink is greater (smaller, respectively) than $p$. Therefore, the rewiring goes with some effective probability  $p_{eff}$. The mean values  of $p_{eff}$ are shown  in Fig. \ref{p_eff}. One should notice that the number of edges rewired each time step is time independent.

\section{\label{sec:Res}Results}
\subsection{\label{sec:Stat}Reaching stationarity}
\begin{figure}
\includegraphics[width=0.48\textwidth]{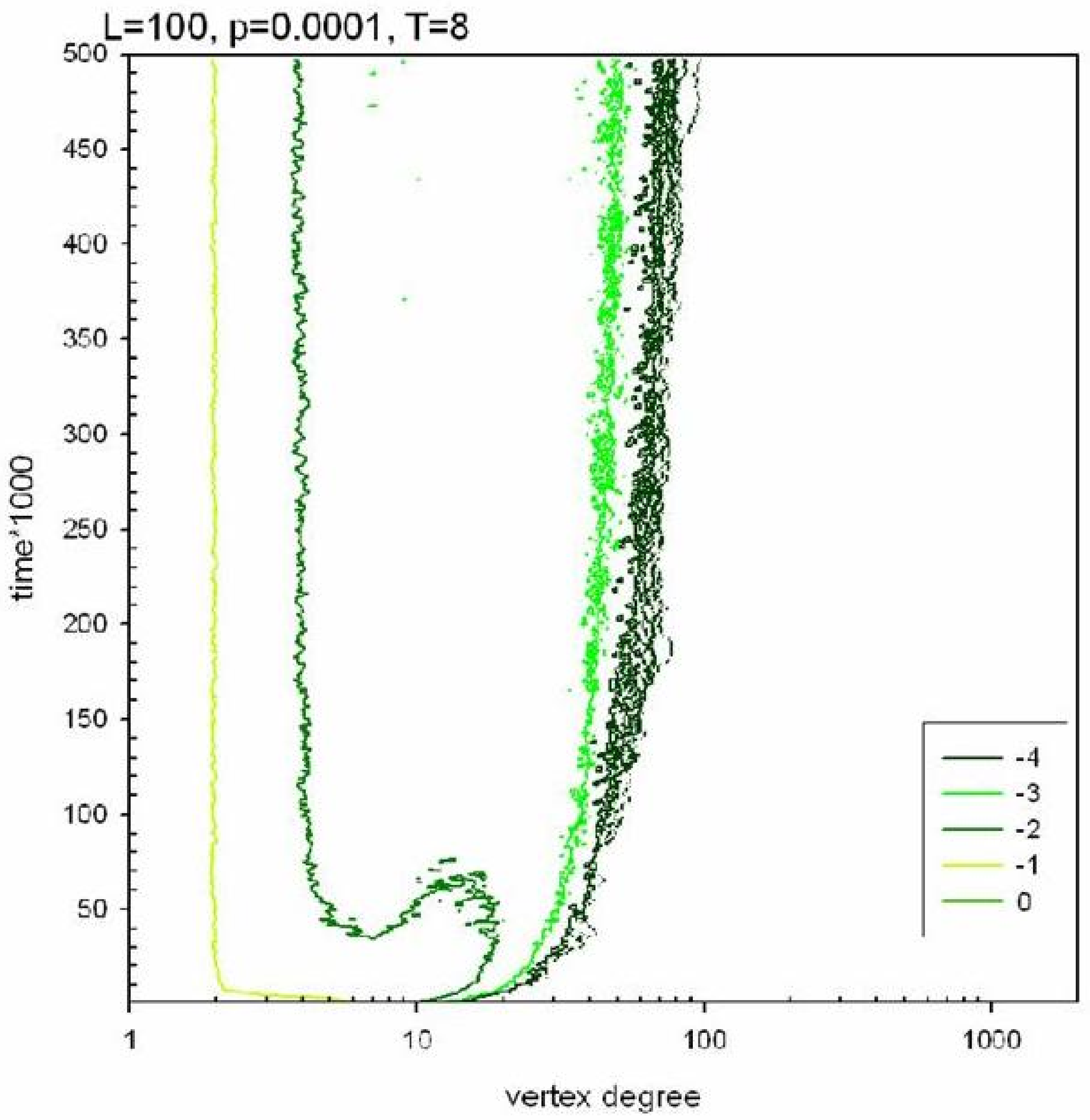}
\includegraphics[width=0.63\textwidth]{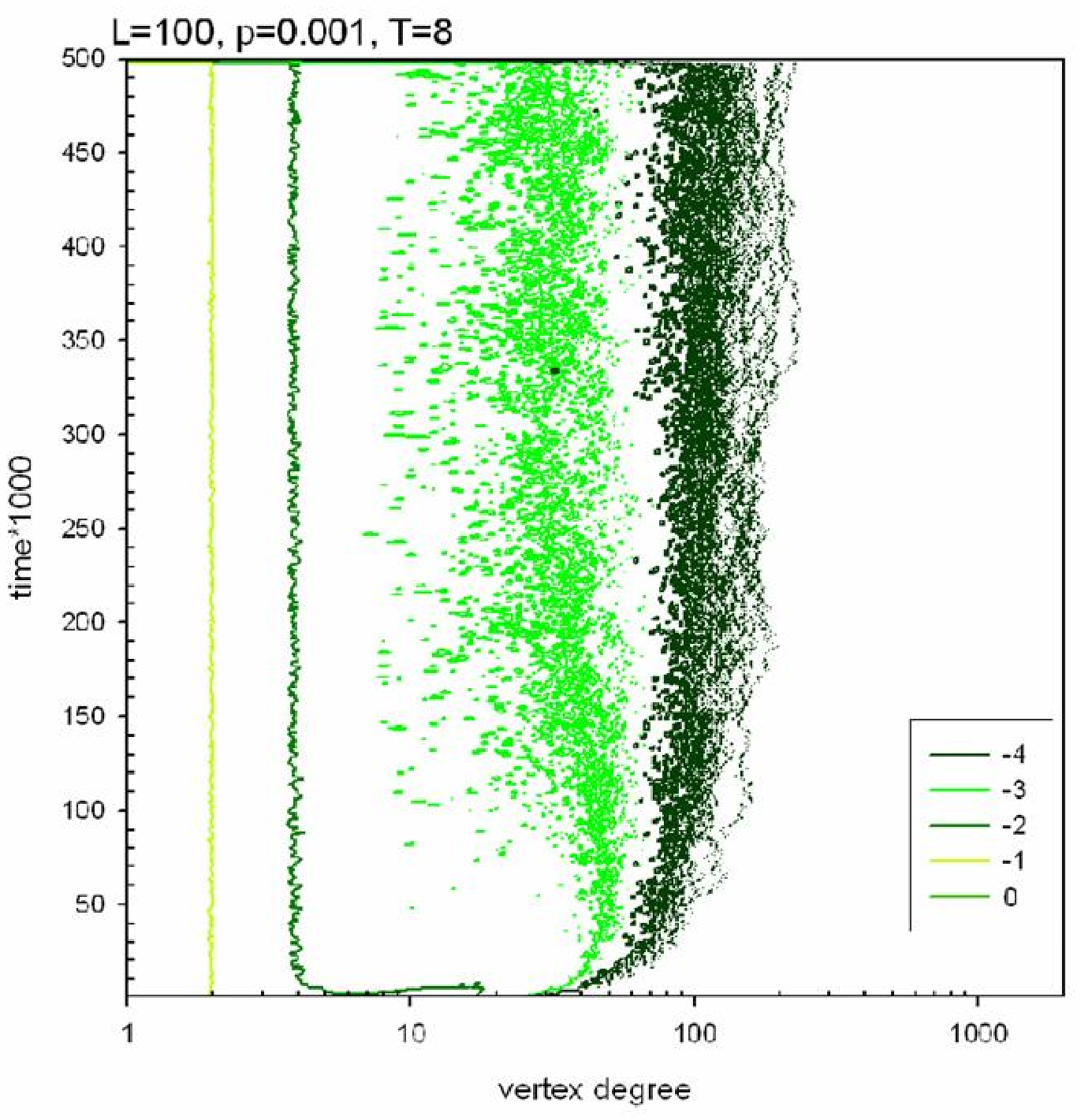}
\includegraphics[width=0.48\textwidth]{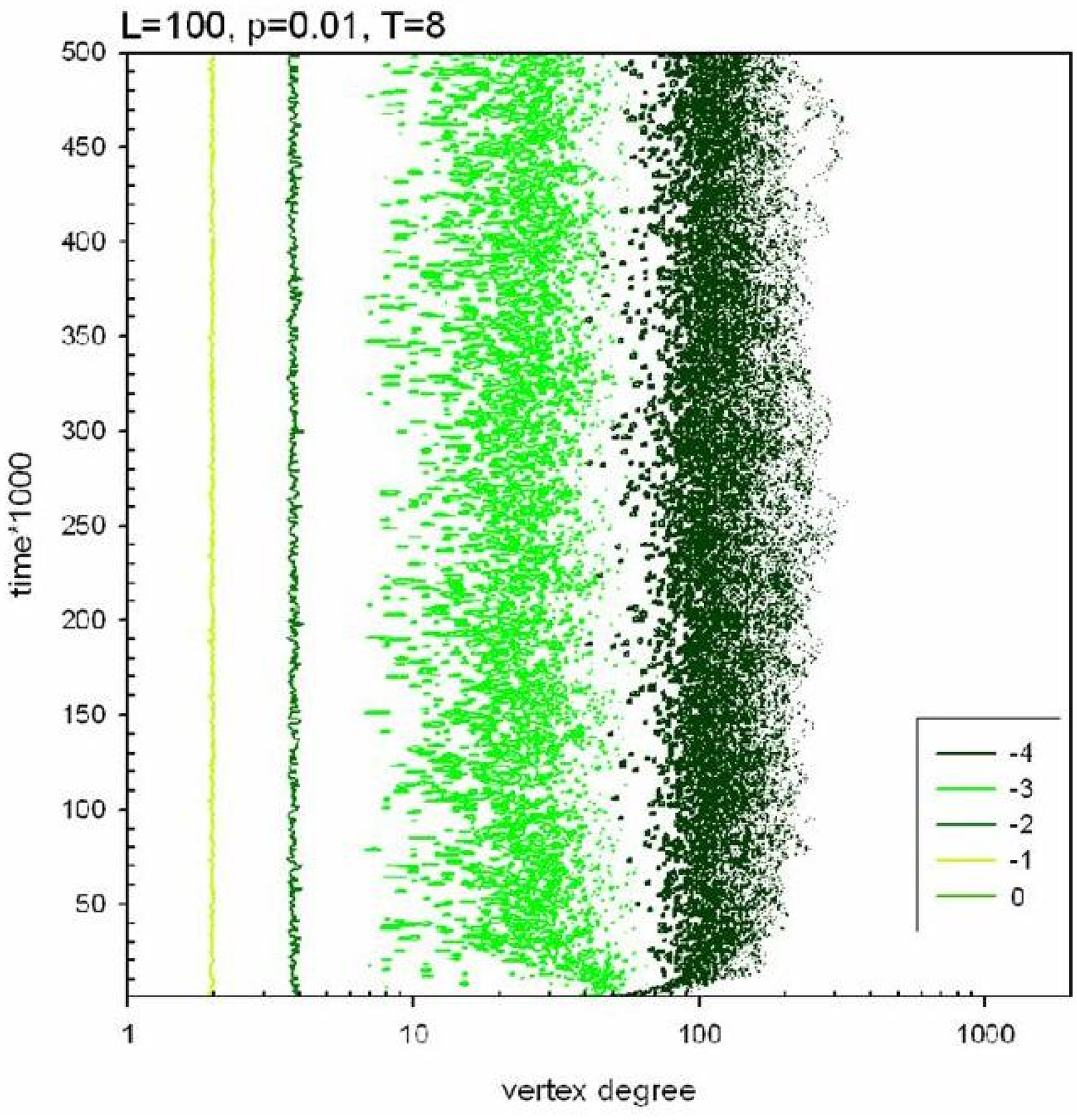}
\caption{Self-organization in the graph structure studied by the distributions of vertex degree in time, case $T=8$, log-scale (color on line). Notice, that  the  graph evolution presented in the top figure can be found as the first $50\,000$ steps of the evolution shown in the middle figure, and the evolution presented in the middle figure can be seen as the first $50\,000$ steps of the bottom figure.}
\label{dist8}
\end{figure}
\begin{figure}
\includegraphics[width=0.48\textwidth]{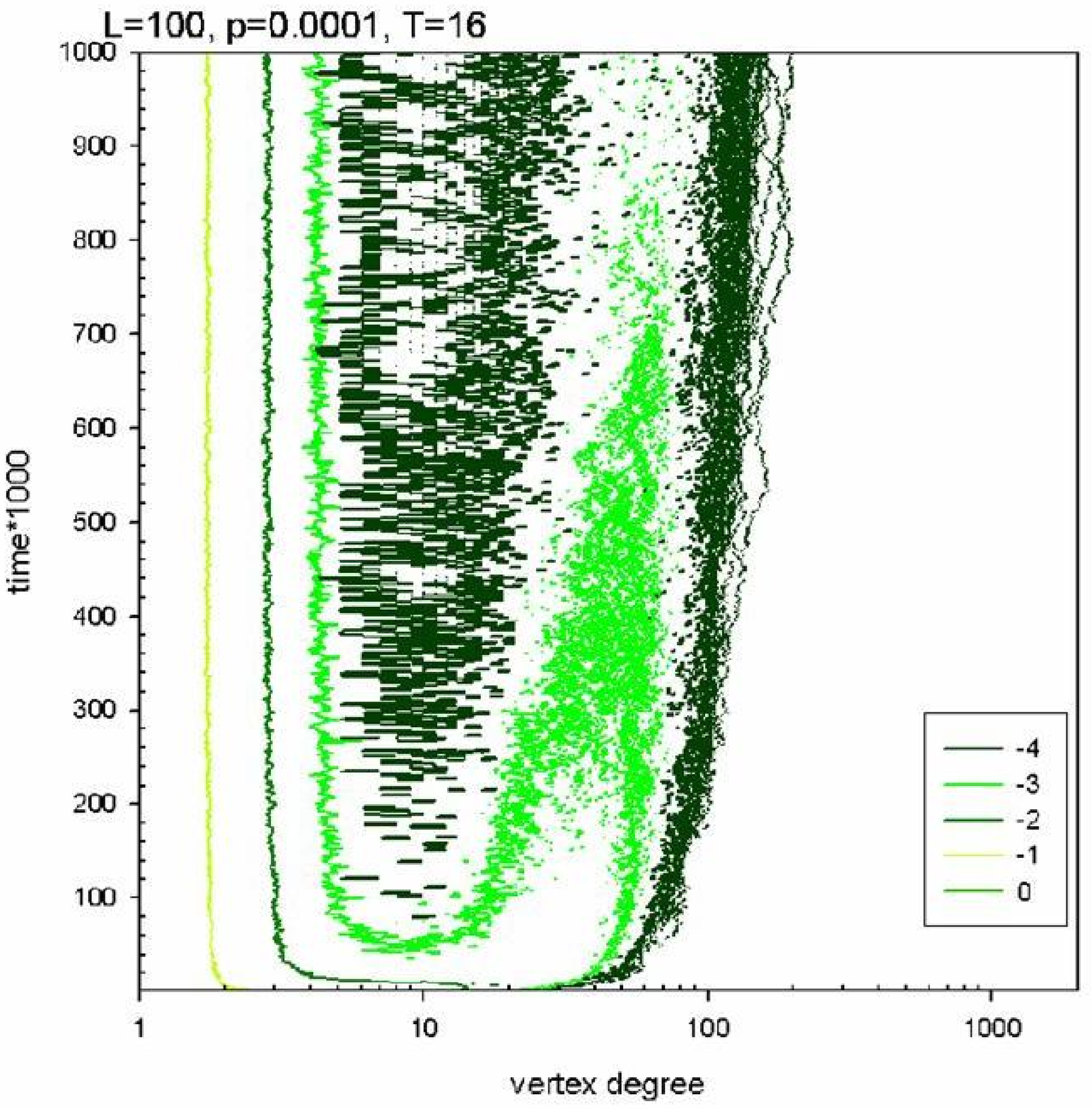}
\includegraphics[width=0.47\textwidth]{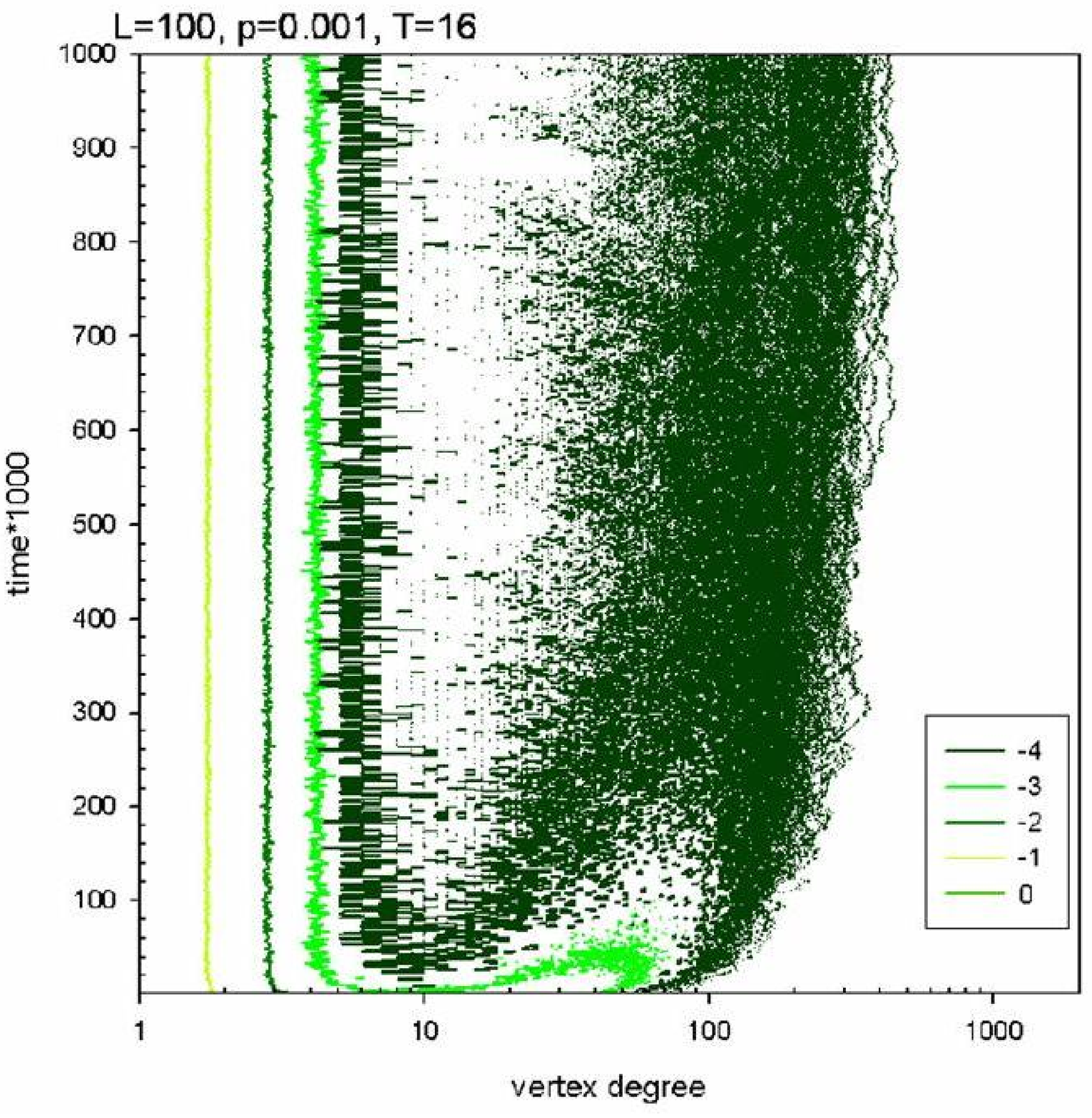}
\includegraphics[width=0.48\textwidth]{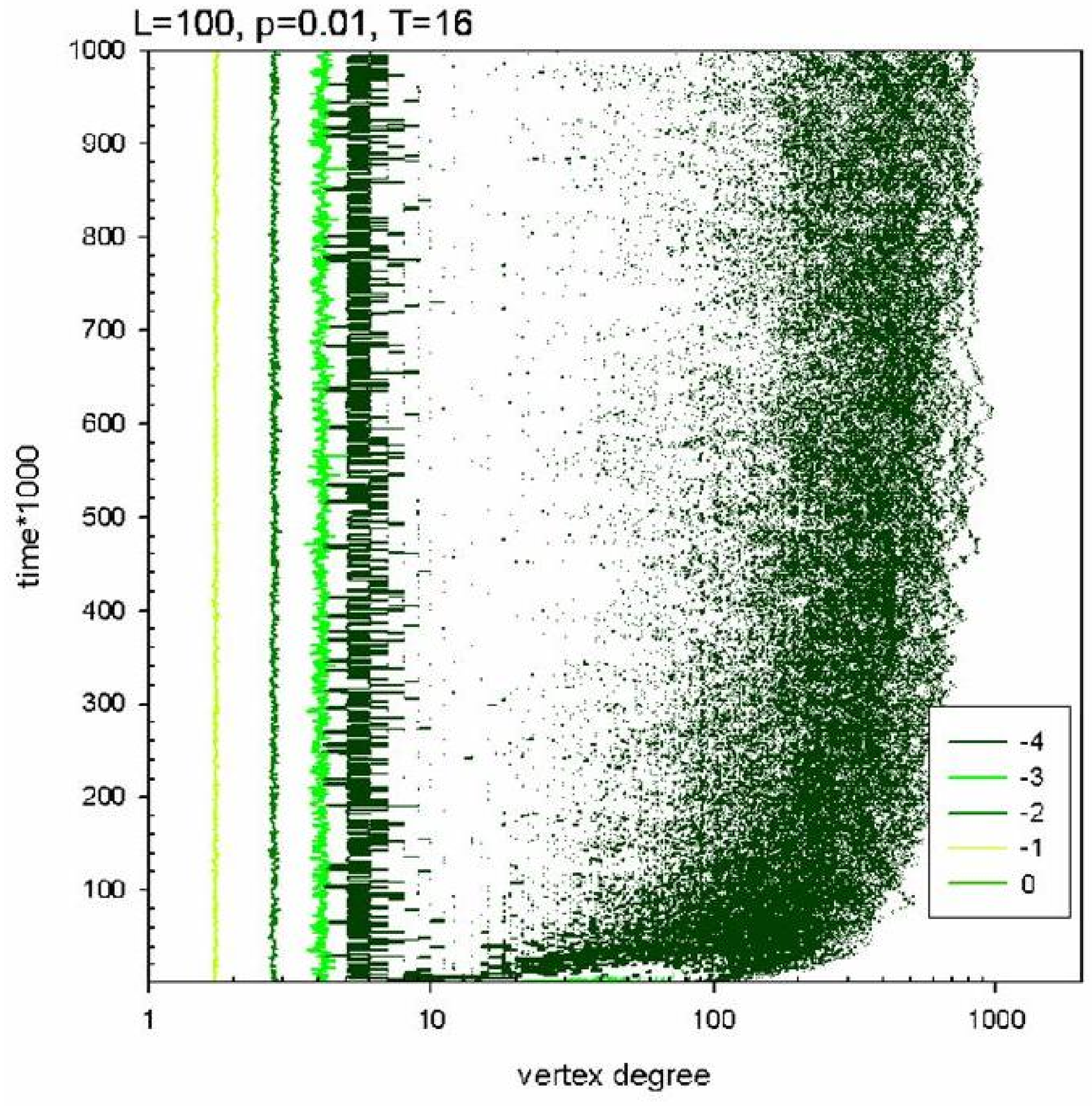}
\caption{Self-organization of the graph structure studied by the distributions of vertex degree in time, case $T=16$, log-scale (color on line). Notice, that  the  graph evolution presented in the top figure can be found as the first $100\,000$ steps of the evolution presented in the middle figure, and the  evolution presented in the middle figure can be considered as the first $100\,000$ steps of the bottom figure.}
\label{dist16}
\end{figure}

We test by simulations the algorithm described in the previous section applied to a square lattice with $L=100$ for different $p$  and with preferences governed by the following two $T$ values $T=8$ and $T=16$. It effects that number of vertices considered is $N=10^4$ and the constant number of edges is $2N$. Since each edge is represented double - on both lists of vertices associated to the edge, the total number of edges considered in rewiring process is $4N$.

\begin{figure}
\includegraphics[width=0.49\textwidth]{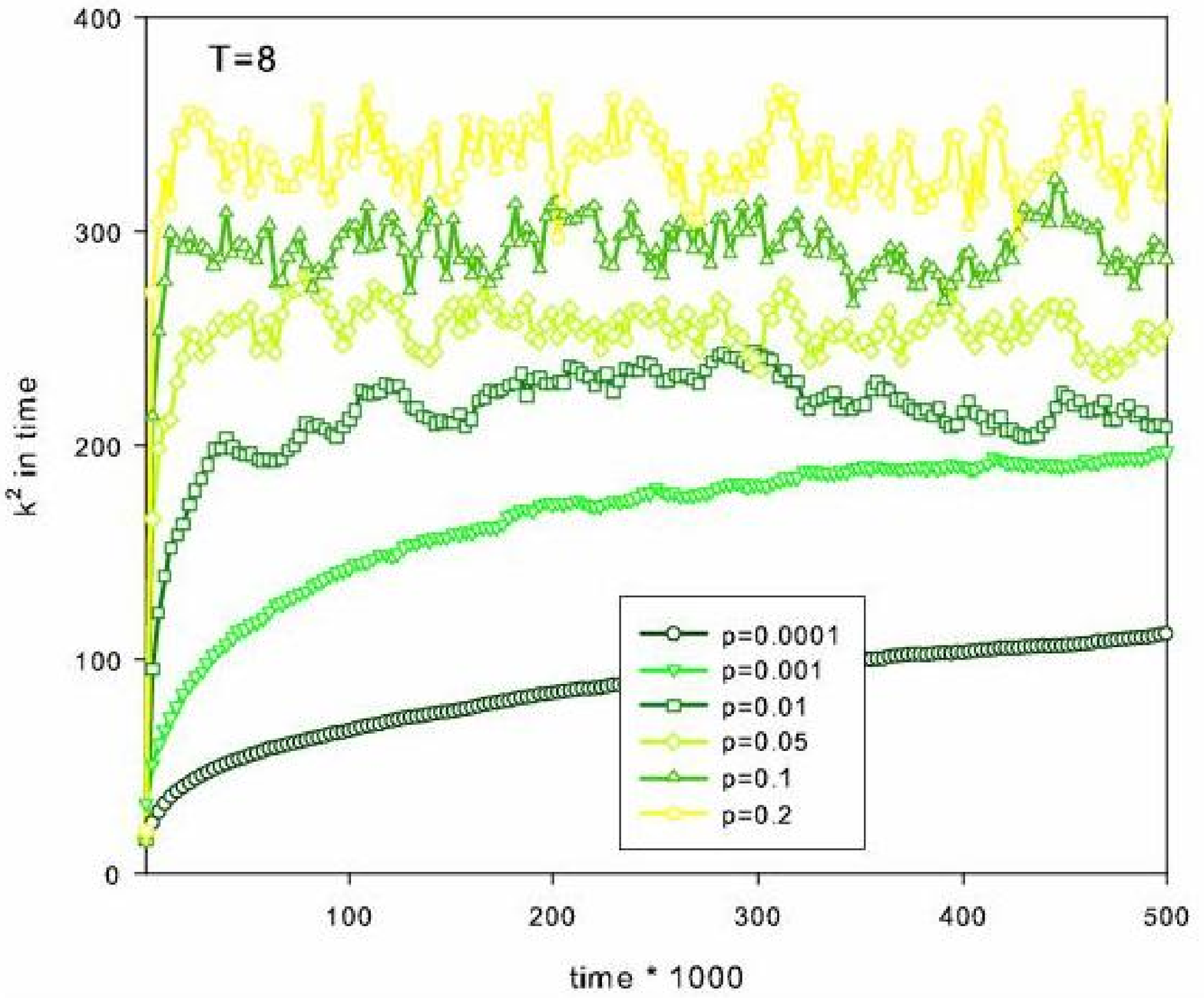}
\includegraphics[width=0.49\textwidth]{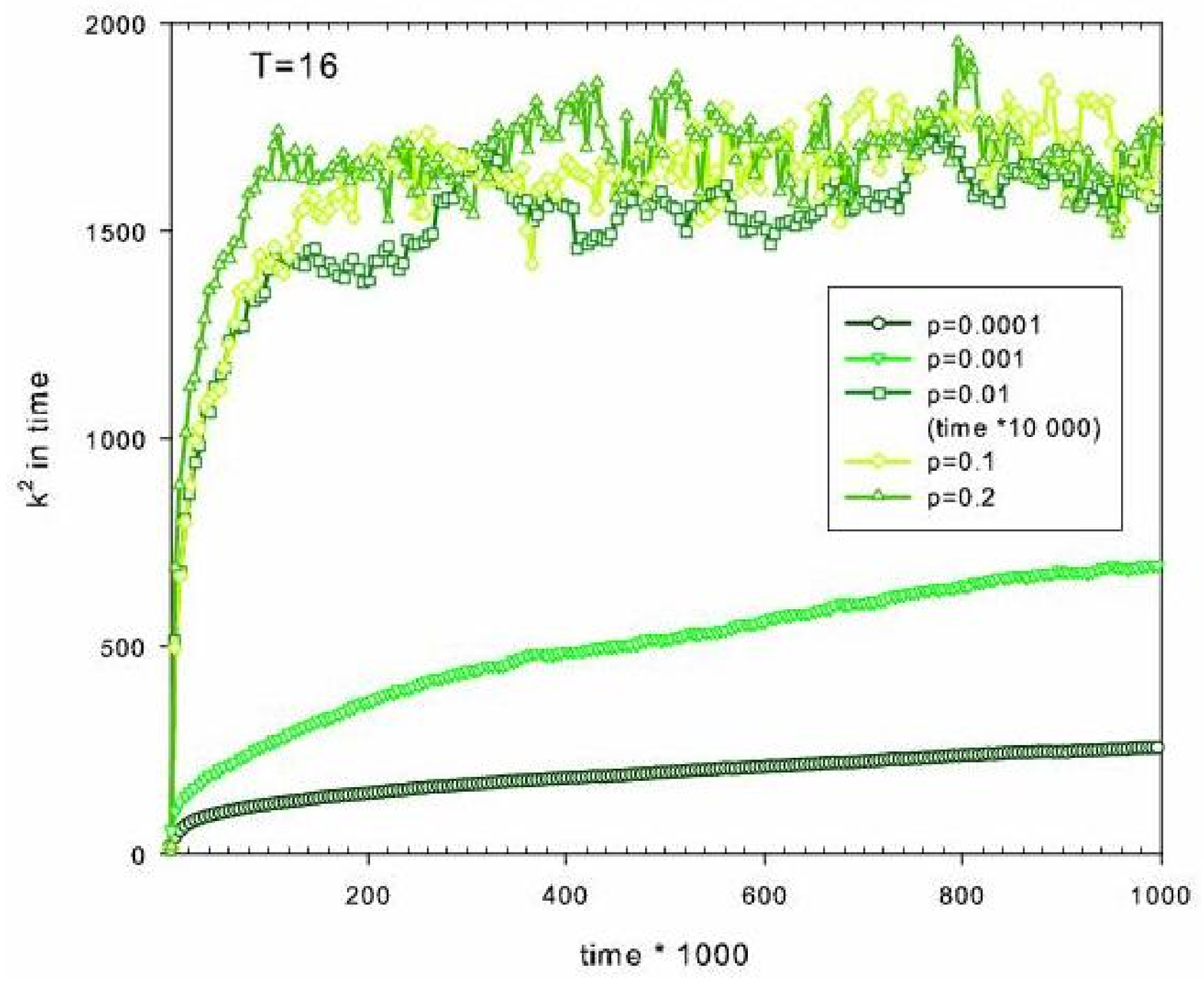}
\caption{Self-organization of the graph structure by second moment of the vertex degree  distributions $k^2$ }
\label{k2time}
\end{figure}

If preferences are not switched on then the network quickly reaches the stationary state with the Poisson degree distribution centered at $k=4$ \cite{Makowiec04}. If the preferences are switched on then for any $p$ and $T$ in a few time steps the initial $\delta$ degree distribution centered at $k=4$ transforms into some other  distribution. Our focus is on the process of  graph restructuring.
In the series of figures: Figs \ref{dist8}, \ref{dist16}, \ref{k2time} we present arguments to prove the fact that a  graph evolving at any $p$ reaches a state for which further edge changes do not influence on the vertex degree distribution. Then we say that the system is self-organized into the fixed stationary state.

In Figs \ref{dist8} and \ref{dist16} typical evolutions of  vertex degree distributions are  presented. The horizontal axis describes vertex degree in log-scale. The vertical axis is for time steps. To quantify properties of the evolving graph, in Fig. \ref{k2time} we characterize the state of the  self-organization  by the second moment of the distributions. The statistical physics of graphs says that graphs with a fixed number of edges compose the canonical ensemble \cite{Dorogovtsev,Farkas,Burda} and the second moment (or equivalently the sum of degrees squared of all vertices) can be interpreted as the energy carried by  a graph with fixed number of edges, \cite{Farkas,Berg}. 

The obtained results show that changes in a graph evolving with $p=0.0001$  are similar to those found for corresponding first steps of a graph evolving with $p=0.001$. The passing time cannot be distinguished from the value of $p$. Comparing figures with $p=0.001$ to figures with $p=0.01$  the similar observation holds. Therefore the first simulation proved  fact is  that $p*4N$ rewirings in each time step after $t$ time steps  is equivalent to the change made by  rewiring $p*t*4N$ edges in one time step for $p$ sufficiently low. In particular, the evolution  with the preference threshold $T=8$ goes in this asynchronous way for all $p \le 0.01$ and the stationary state of the asynchronous evolution is reached within the observed $ 500\, 000$. If $T=16$ then the stabilization of a system needs more time. Therefore all our observations are of $10^6$ time steps long. Nevertheless to reach the asynchronous dynamics ensemble in case of $p=0.01$ we let the system evolve ten times longer, so that we can state again that for all $p<0.01$ the same stationary state is observable, see legends to Fig. \ref{k2time}.

With increasing $p$, namely $p\ge 0.2$, stationary states are reached definitely  faster for both $T$ values. Here the `synchronized' preferential rewiring dominates in one-step transition what results that at each $p$ the system constitutes different stationary ensemble. These ensembles  are distinct quantitatively from each other by the different vertex degree distributions what reflects in the distinct values of $k^2$.

\subsection{\label{sec:dist}Stationary state vertex distribution}   
\begin{figure}
\includegraphics[width=0.49\textwidth]{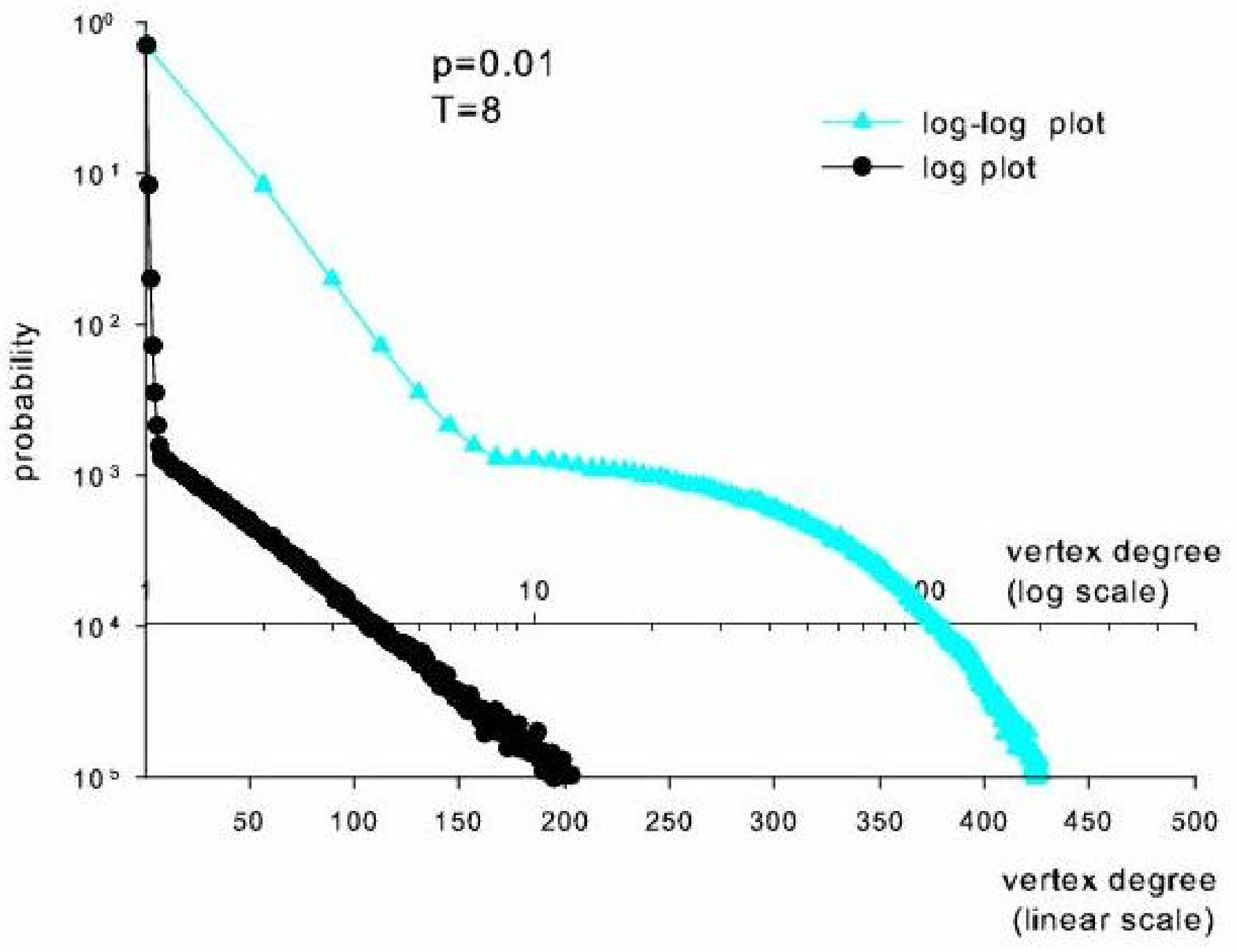}
\includegraphics[width=0.47\textwidth]{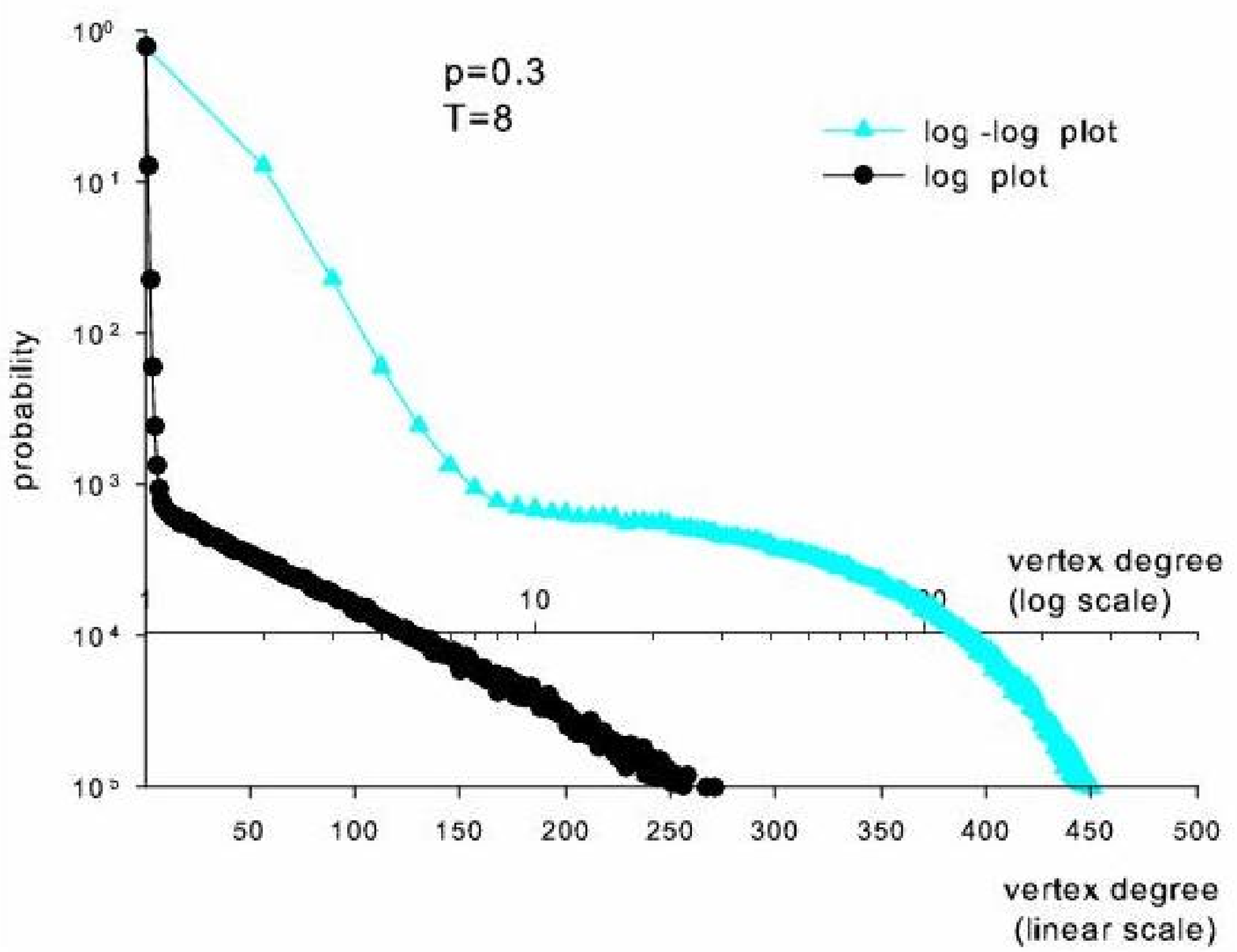}
\includegraphics[width=0.49\textwidth]{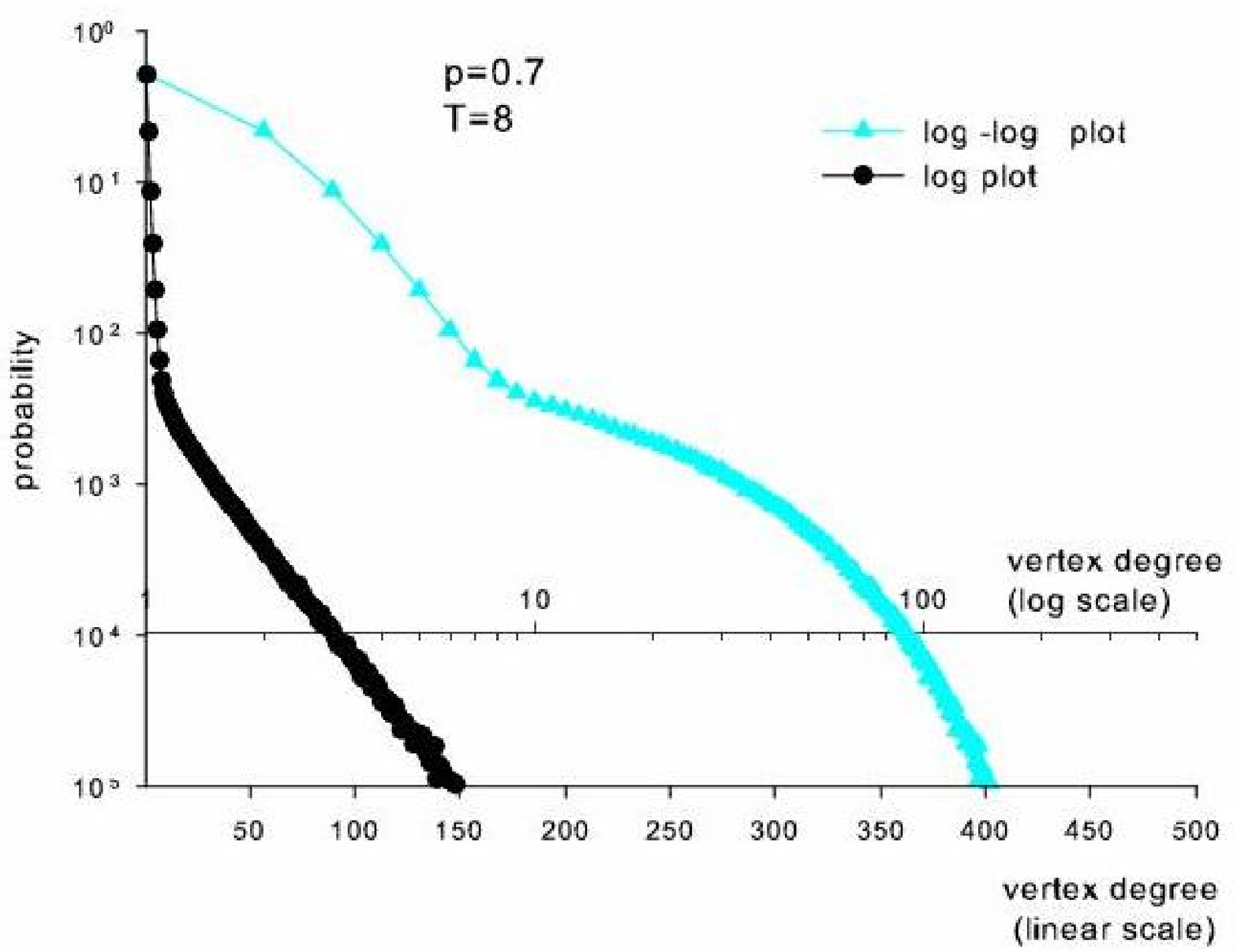}
\caption{\label{T8a}The stationary vertex  degree distributions in case $T=8$ and at different role of synchronous dynamics: the asynchronous dynamics results (top figure) and stationary distributions  resulting from different   synchronous evolution rule.  The two horizontal scales are used to amplify properties. The ordinary linear scale (bottom) is  to observe exponential dependences and the log scale (black plots) is to show the power-law degree regions of decay (gray plots). (color on line).}

\end{figure}

\begin{figure}
\includegraphics[width=0.49\textwidth]{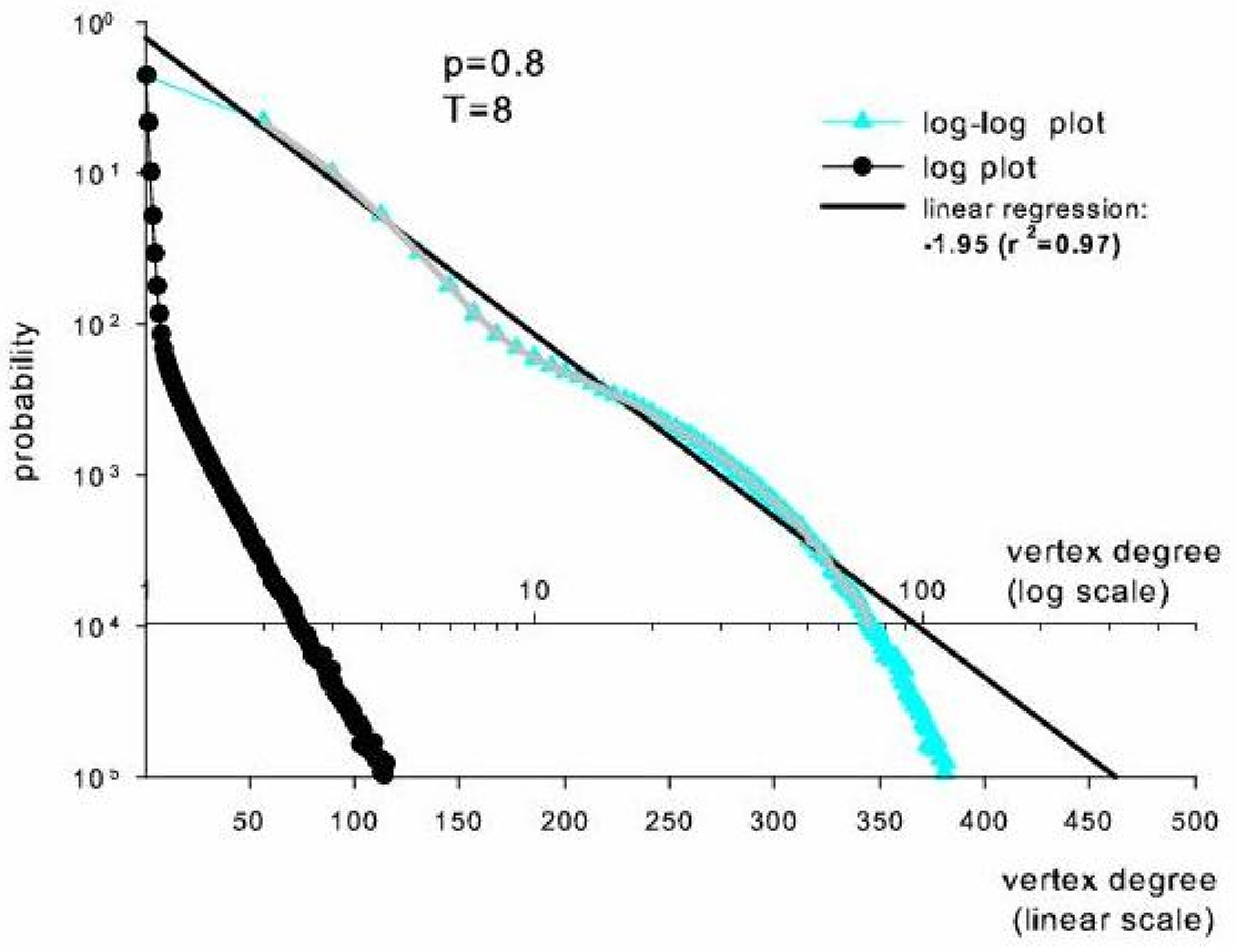}
\includegraphics[width=0.49\textwidth]{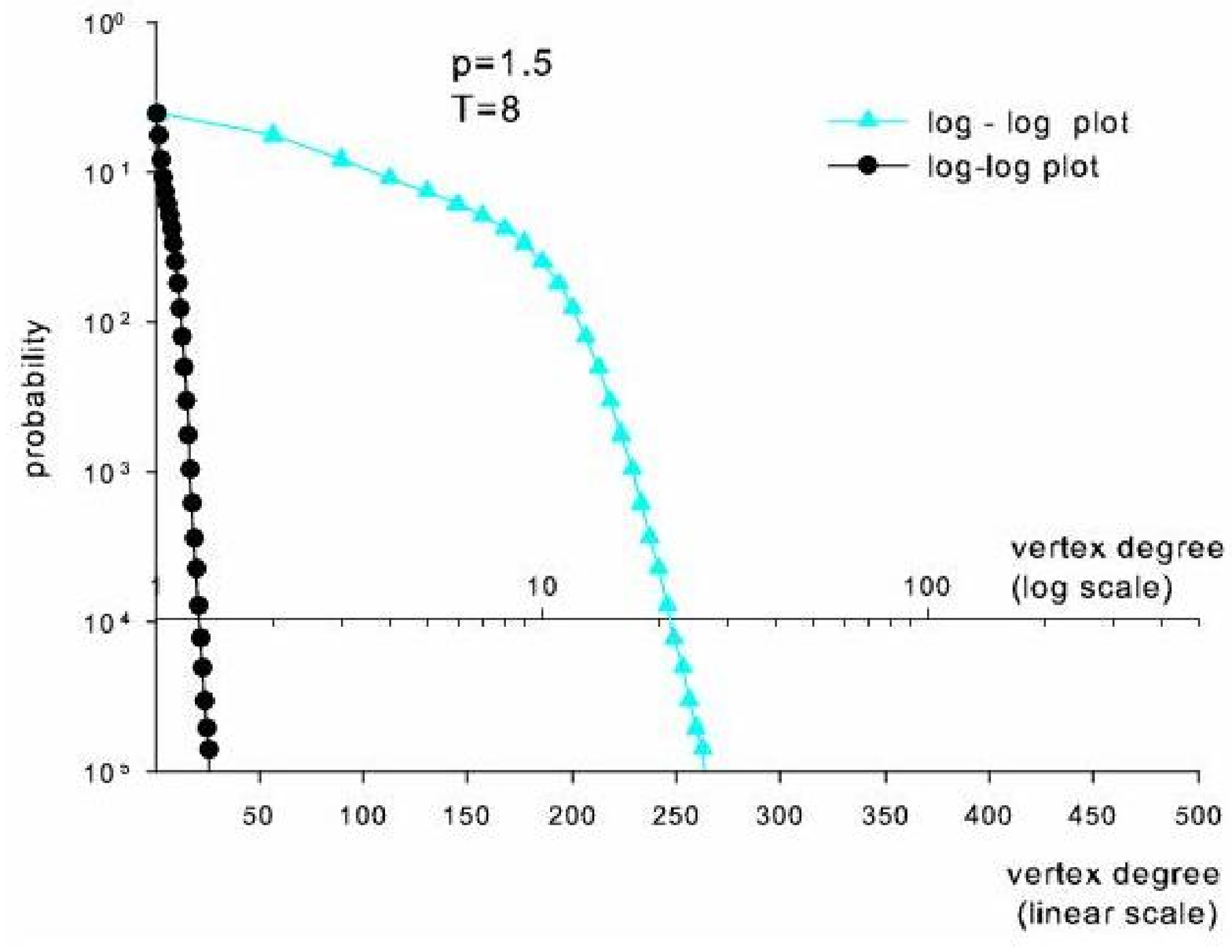}
\caption{\label{T8b}The stationary  vertex degree distribution in case  $T=8$ and $p$ when the largest interval of vertex degrees is of a power-law type (upper figure, the gray plot) and the stationary distribution of a vertex degree in case of the most  synchronous dynamics (bottom figure). The two horizontal scales are used to amplify properties. The ordinary linear scale (bottom) is  to observe exponential dependences and the log scale (black plots) is to show the power-law degree regions of decay (gray plots). (color on line).}
\end{figure}

\begin{figure}
\includegraphics[width=0.49\textwidth]{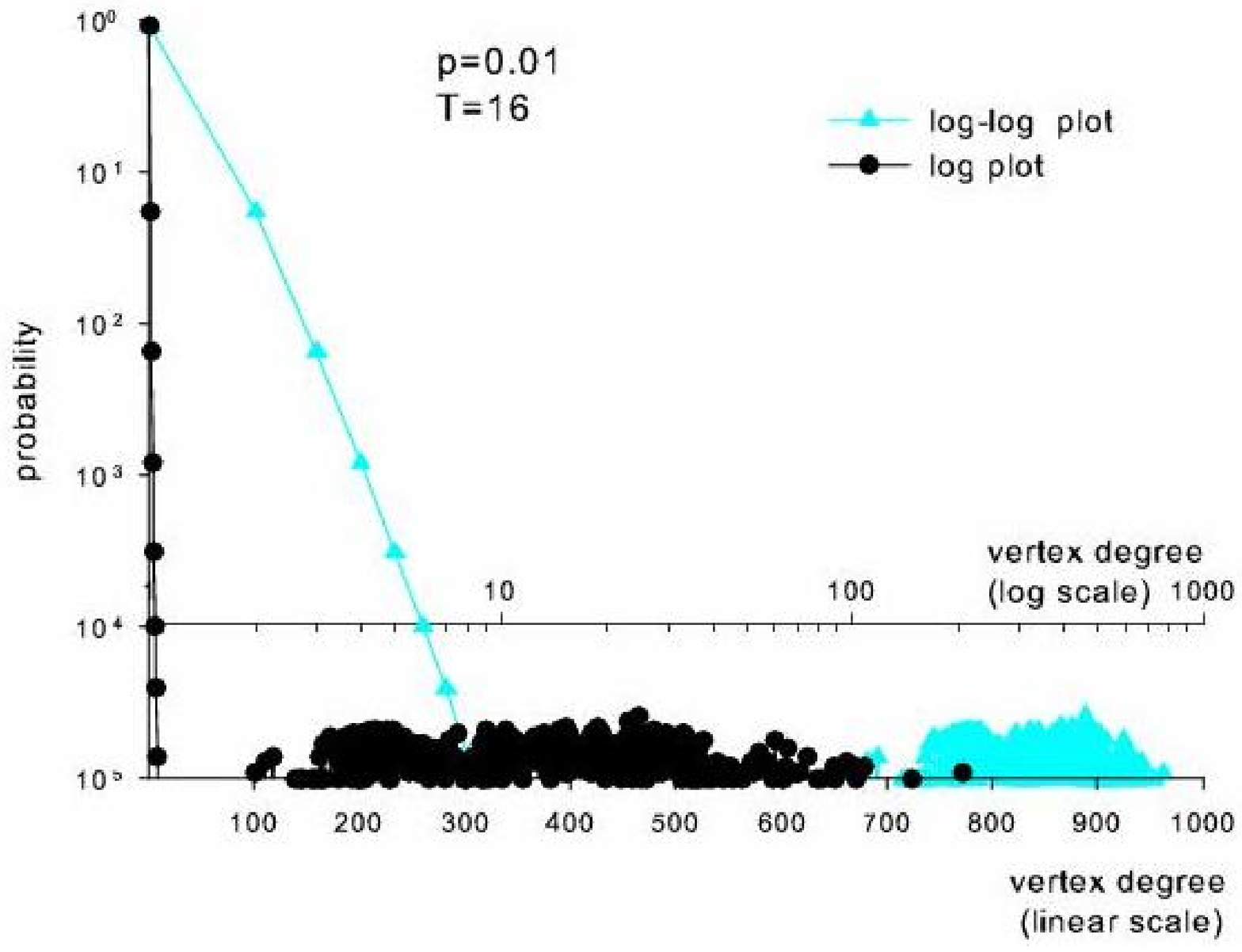}
\includegraphics[width=0.47\textwidth]{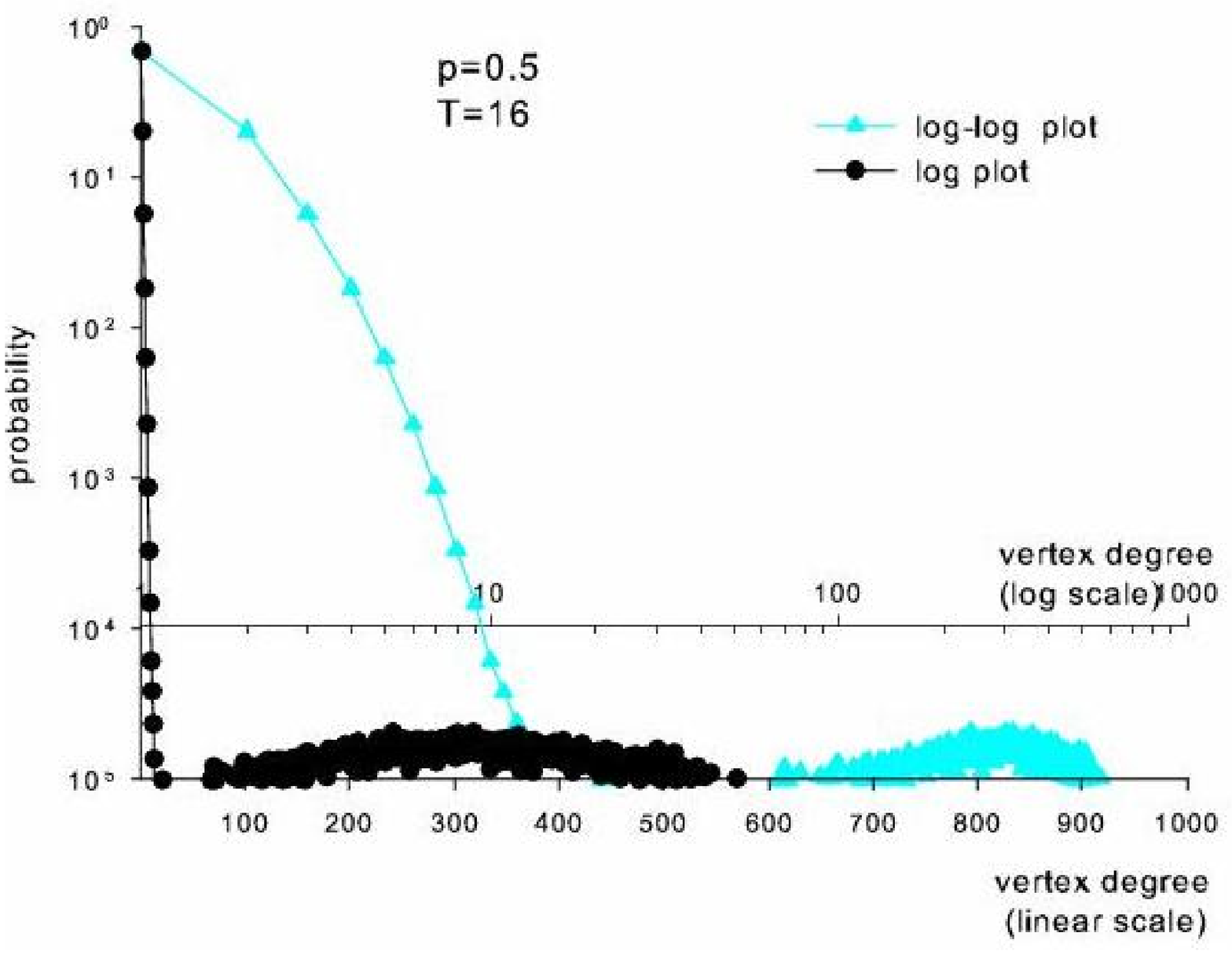}
\includegraphics[width=0.49\textwidth]{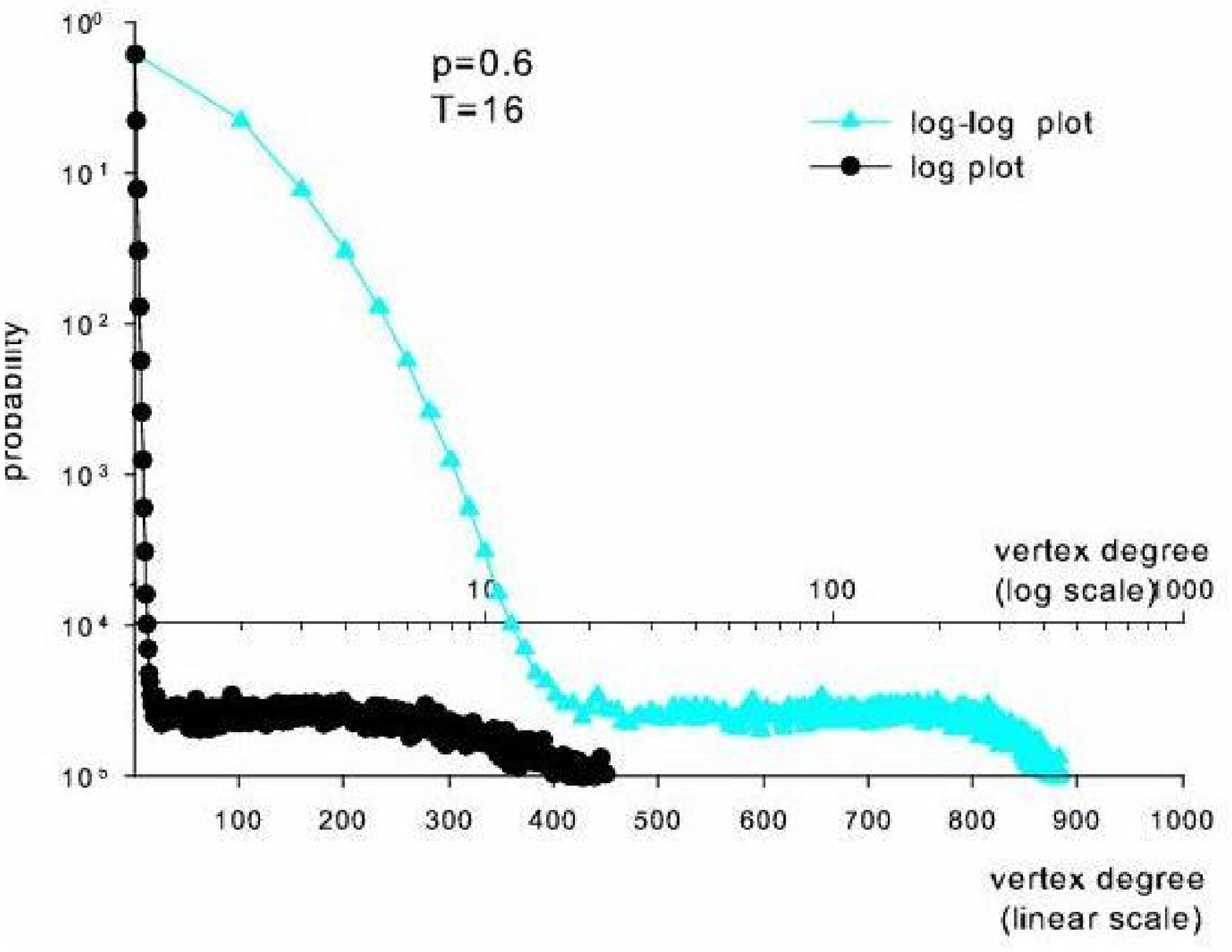}
\caption{\label{T16a}The stationary  vertex degree distributions in case  $T=16$ at different role of synchronous dynamics: the asynchronous dynamics results (top figure) and stationary distributions  resulting from different   synchronous evolution rule. The two horizontal scales are used to amplify properties. The ordinary linear scale (bottom) is  to observe exponential dependences and the log scale (black plots) is to show the power-law degree regions of decay (gray plots). (color on line).}
\end{figure}

\begin{figure}
\includegraphics[width=0.49\textwidth]{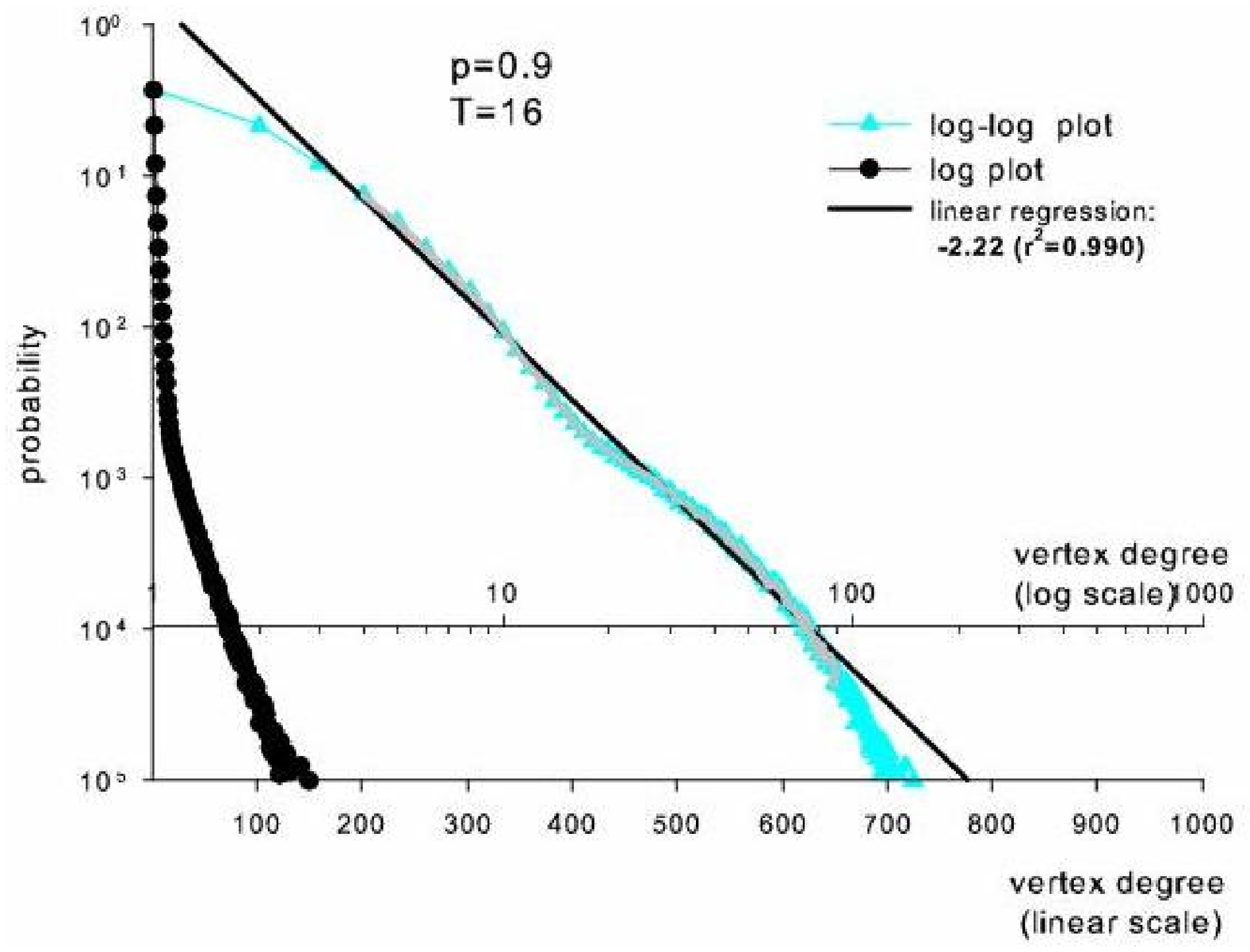}
\includegraphics[width=0.49\textwidth]{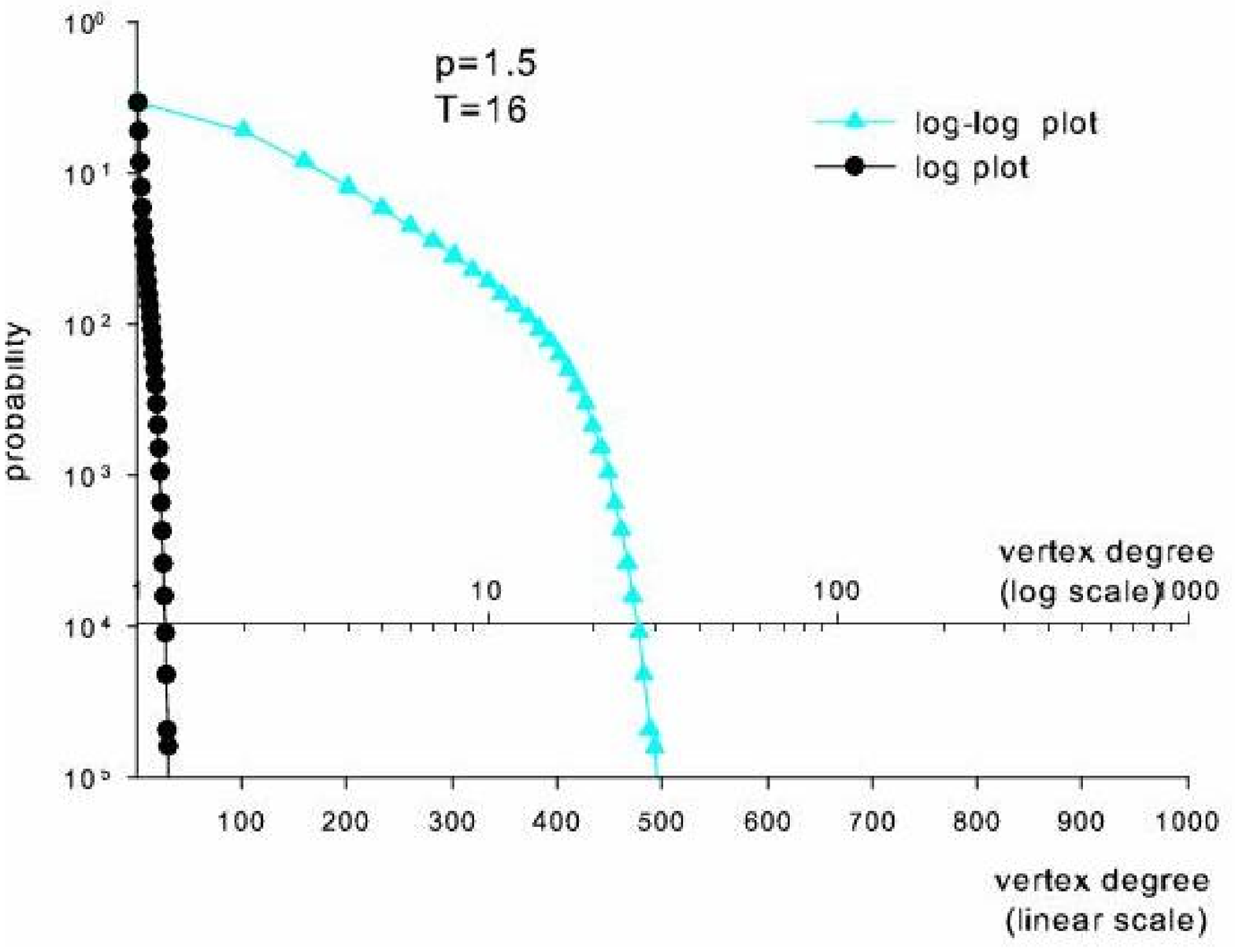}
\caption{\label{T16b}The stationary  vertex degree distribution in case  $T=16$  and $p$ when the largest interval of vertex degrees is of a power-law type (upper figure, the gray plot) and the stationary distribution of a vertex degree in case of the most  synchronous dynamics (bottom figure). The two horizontal scales are used to amplify properties. The ordinary linear scale (bottom) is  to observe exponential dependences and the log scale (black plots) is to show the power-law degree regions of decay (gray plots). (color on line).}
\end{figure}

In Figs \ref{T8a}, \ref{T8b}, \ref{T16a} and \ref{T16b} we show distributions of  the stationary states obtained at different $p$ and for $T=8$ and $T=16$, respectively. 


In Fig. \ref{T8a} the top panel describes the graph ensemble resulting from asynchronous evolution. The degree distribution is like a  two-part function. The first part represents properties of vertices which are not preferred, i.e., vertices with its degree smaller than the threshold ($k < 8$).  The second part of the distribution represents the tail properties, i.e., probabilities to find vertices with  degrees much greater than the threshold $T=8$, $k >> T$. The tail decays exponentially with the rate $\alpha= 0.016 (0.99)$ (the Pearson $r^2$-error is added in brackets). Between these two separated regions there is a  transient region, here  the transient region means $8 \le k\le 40$, which can be described by the power-law dependence on vertex degree. The exponent for this decay is $\gamma=0.43 (0.97) $. Notice, that this transient region is not visible on the exponential curve.

With rewiring rate $p$ increasing, the exponential tail decay slows down (at $p \in (0.3,0.4) $,  $\alpha=0.0068 (0.99)$ and this is the minimal value observed  by us) and then grows up to $\alpha =0.017(0.95)$ at $p=0.7$, see the second and third panels in Fig.\ref{T8a}. Together with fast tail decay, the transient region with power-law characteristics links to the first part of the distribution  what forms  the largest vertex  degree interval with power-law type of distribution. At $p=0.8$ we have the exponent of the power-law decay $\gamma=1.95 (0.97)$ for $2\le k\le 70$, see Fig. \ref{T8b} top panel. However, for $k>70$ the decay changes into $\gamma=6.6 (0.96)$. Finally, when $p\ge 1$ then the transient region vanishes. The decay corresponding to vertices with $k\le 8$ can be still approximated by a power-law but then if $k>8$ the fast exponential decay occurs. In particular at $p=1.5$ we have  exponential rate decay $\alpha=0.215(0.99)$ and power law decay with exponent $\gamma=0.86 (0.96)$ when $k \le8$, see Fig. \ref{T8b} bottom panel.

In case $T=16$ the first and  tail parts of degree distribution are isolated from each other for a large interval of $p$, see the first and second panels in Fig.\ref{T16a}. The noticeable transient region constitutes at $p=0.6$ ( $\gamma \approx 0 $ for $20\le k\le 360$ ), see Fig.\ref{T16a} the bottom panel. With growing $p$, similarly to the ensembles resulting from evolution with $T=8$, we observe  the junction  between the first and transient regions of the distribution, and the largest interval of the  power-law decay establishes. Namely, when $p=0.9$ then for $ 2\le k\le 80$ $\gamma = 2.22 (0.99)$, see the top panel in Fig.\ref{T16b}. However for vertices with a degree $k>80$ we observe the fast exponential-like  decay. Stationary ensembles arising from the evolution with $p\ge1$ are similar to those described in case $T=8$. The first part, which can be approximated by power-law, spreads to $k=16$. Then the fast exponential decay occurs, see Fig.\ref{T16b}.

\subsection{\label{sec:trans}Phase transition}   
\begin{figure}
\includegraphics[width=0.49\textwidth]{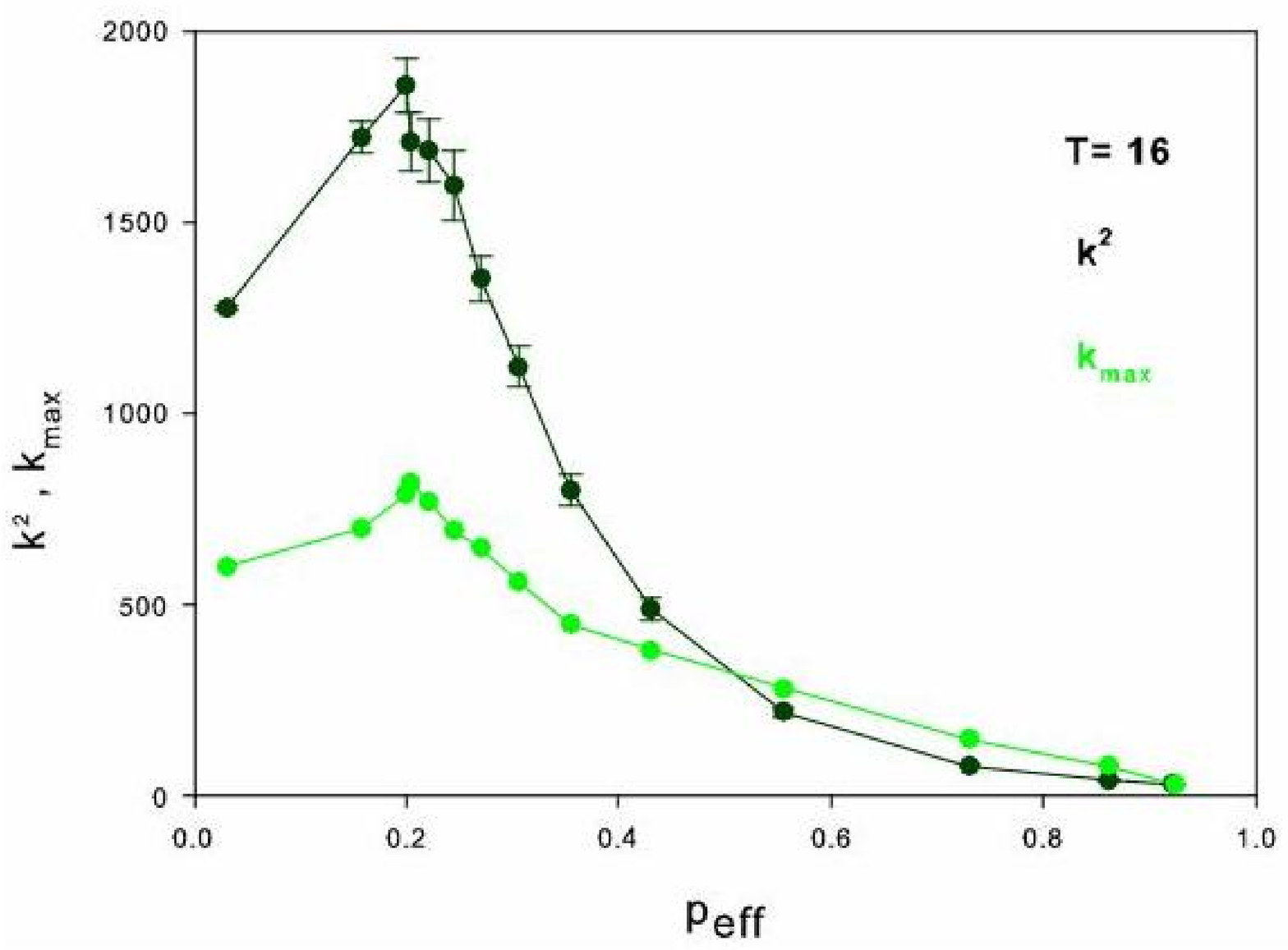}
\includegraphics[width=0.49\textwidth]{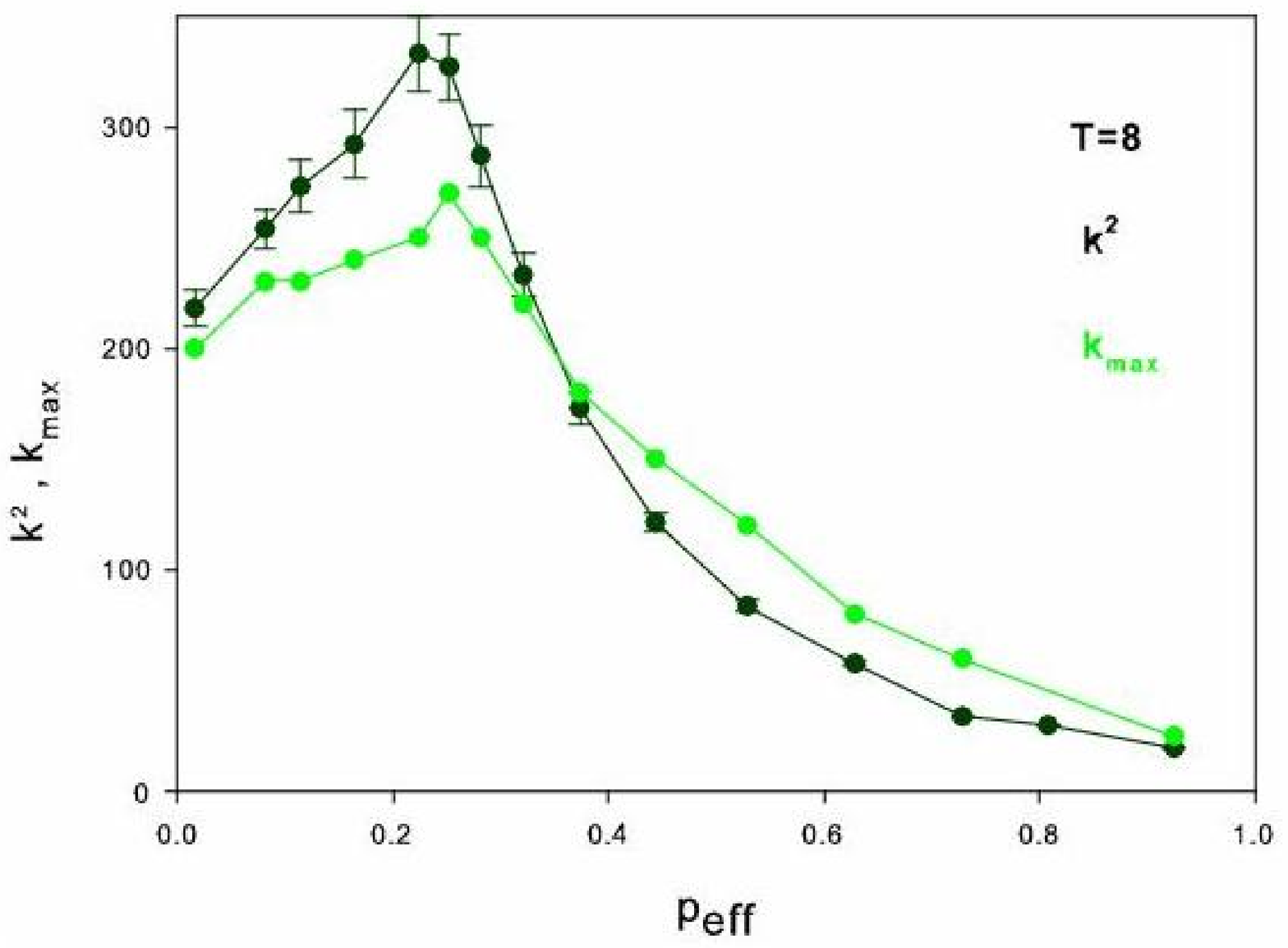}
\caption{ The second moment of the vertex degree distributions (with standard deviation errors) and maximal vertex degree (probability that such a vertex occur on a lattice is greater than 0.1) in the stationary regime, (color on line).}
\label{k2}
\end{figure}

From the vertex degree distribution study it appears that at fixed  $T$ there exist two
different self-organized graph ensembles corresponding to the small $p$ and large $p$ value. It is said that if some global statistical property representative for a graph topology changes then such a phenomenon is referred to as topological phase transition \cite{Farkas}. As the most appropriate order parameters to describe this transition we consider $k^2$ ---  mean energy in the ensemble, and $k_{max}$ --- the largest vertex degree which occur with probability greater than $10^{-5}$. Fig. \ref{k2} show dependences of both order parameters on $p_{eff}$. For  both $T$ values and both order parameters two regions of $p_{eff}$ are observed. The point of topological phase transition  can be localized as $p_{eff}^{crit} \in (0.2, 0.3) $

The two phases in graph topology can be described as follows: 

{\noindent}{\it Leafy  phase}: \\
There is a huge component consisting of almost all vertices and few little 2, 3-vertices graphs. Small number of vertices with extremely high degrees (hubs) serve as centers of this component. These centers are  multiple interconnected between each other what together establishes a firm long living graph skeleton. The skeleton is close to the complete graph.  Hence the strong assortativity between hubs is present. Since the vertices of the graph skeleton are surrounded by plenty of leaves then one can find the strong disassortativity between skeleton vertices and leaves. At low $p$ about 90\% of vertices are leaves and therefore  we call a typical network of the ensemble as leafy-complete graph and the corresponding phase as leafy phase. 

{\noindent}{\it Tangling phase:} \\
At $p\ge 1$ the strong expectation can arise  that the network ensemble should collapse to a stochastic graph. This expectation is supported  by the fact that when $p \ge 1$ then there exists a degree interval, namely $T\le k \le pT$  where the probability to unlink and probability to link to are equal to $1$ - like  in a stochastic graph. Moreover, vertices with $k > pT$ are still of high probability to be unlinked, see Fig.\ref{prefy}.
The stochastic  graph distribution is distinguished from others by the peaked shape around the average degree value, here  sharp maximum at $k=4$ should appear. However, such distribution did not appear at any model parameters considered by us. The dynamics which persistently   applies the preference rule, drives the system to other than a stochastic graph solution. There is not a typical vertex for network. Instead  vertices $k\le T$ occur with similar probability. One can think that each time step plenty of edges are  moved around between the similar low degree vertices what blocks the possibility to establish vertices with higher degrees.  Let us call this network ensemble as tangling net to underline activity and inefficiency of the evolution rule. The corresponding phase will be called tangling phase.

The process of transition from the leafy to tangling phase can be explained qualitatively in the following way. In leafy phase there is a pool of edges attached to vertices of little degree and a small number of vertices with  high degrees which are densely interconnected between each other. The presence of multiple connections makes  the skeleton connections degenerated what results in that the effective number of edges in the system is much smaller than $2N$. With growing $p$ the information about changes in the edge connections are not immediately available to other rewirings what  weakens  the assortativity between hubs.  Therefore the graph skeleton obtains less possibility to emerge and  the mechanism analogous to the preferential attachment in growing models of scale free networks \cite{BA99} can occur. `New' edges appear due to vanishing connection degeneracy. Though the appearance of the graph skeleton is less probable, but centers which accumulate edges occur due to the persistently applied dynamics with preferences.

\section{\label{sec:Sum}Conclusions and discussion}
By considering the network evolution as two-step: the preferential rewiring of edges and updating of the information about changes done we obtain a discrete time network evolution. In this time step scale the evolution drives a regular lattice to the graph ensemble with fixed statistical properties. The presented simulation foundings can be summarized as follows:
\begin{enumerate}
\item Networks evolving with preferential rewiring rule self-organize into stationary states. 
\item The stationary states form ensembles of exponential-type networks. 
\item The statistical properties of the stationary states such as vertex degree distribution, its second moment and maximal vertex degree are related to model parameters $p$ and $T$.
\item In the space of stationary states two separated regions are identified:\\
--- the limit of asynchronous preferential evolution, called leafy phase, if $p\le 0.01$ and  $T=8, 16$ \\
--- the limit of synchronous higly rewired preferential evolution, called tangling phase, if $p > 1.0$ and $T=8, 16$\\
Between these two phases for different model parameters the different final states occur.
\item The maxima of the second moment of vertex degree distribution and the largest vertex degree are localized at $p\approx 0.2$ for both $T$. This feature suggests that the transition point between the two phases observed can be related to this value.
\item The transition goes by smashing multi-connections between vertices belonging to the graph skeleton. Since the wide interval with the power-law decay in the degree distribution  is observed we can claim that the mechanizm which is characteristic for Barabasi-Albert network,  namely the growth in the network \cite{BA99}, is present. But here this mechanizm denotes the  increase in the  number of effective  egdes. 

\end{enumerate}

The two most important physical mechanisms are usually listed to explain the occurrence of power-laws \cite{Newman05}. (1) {\it a rich-get-richer} mechanism in which the most linked vertices get more links. Here such  mechanizm is represented by the preferential attachment rule and moreover this mechanizm introduces a kind of ordering. (2){\it critical phenomena} where the ordering mechanizm is in conflict with some dissordering process. Here, a temperature like mechanizm can be ralated to the synchronization. The synchronization weakens the order coused by the preferential rule. This confict effects in appearence of critical  properties.

Rough investigations about the graph ensembles arisen when $T=100$ have been done but only for $p \ge 0.1$ to limit the length of runs. Again, the limit stationary degree distributions can be divided into two classes: the leafy-complete graph ( $p=0.1$ ) and the exponential graph ( $p=1.0$ ). However, for any value of $p$, there was not noticed a power-law dependence in the first part of the distribution as well as there was not observed any transient region. Hence, in the phase-space of $T$ parameter another  topological transition can be localized.

\noindent{\bf Acknowledgment }\\
We wish to acknowledge the support of Polish Ministry of Science and Information Technology Project: PB$\slash$1472$\slash$PO3$\slash$2003$\slash$25

\end{document}